\documentclass[aps,prl,twocolumn,floatfix,superscriptaddress]{revtex4-1}
\usepackage{amsfonts,amsmath,amsthm,amssymb,amsopn,graphicx}
\usepackage{dsfont,bbm,xcolor}
\usepackage{enumitem}
\usepackage[hyperindex,breaklinks]{hyperref}
\usepackage{breqn}
\usepackage{siunitx}
\usepackage{color}
\makeatletter
\let\cat@comma@active\@empty
\makeatother

\newcommand{\ii}{ {\rm i} }
\def\bra#1{\mathinner{\langle{#1}|}}
\def\ket#1{\mathinner{|{#1}\rangle}}

\newcommand{\ave}[1]{{\langle #1\rangle}}
\newcommand{\z}{{\rm z}}

\newcommand{\tr}{\rm{tr}}

\def\one{\mathbbm{1}}
\def\bra#1{\mathinner{\langle{#1}|}}
\def\ket#1{\mathinner{|{#1}\rangle}}

\newcommand{\dd}{ {\rm d} }

\newcommand{\x}{{\rm x}}

\newcommand{\LL}{{\hat{\cal L}}}

\def\tr{{{\rm tr}}}

\def\one{\mathbbm{1}}
\def\Re{{\,{\rm Re}\,}}
\def\im{{\,{\rm Im}\,}}

\def\Re{{\,{\rm Re}\,}}
\def\im{{\,{\rm Im}\,}}

\begin{document}
\title{Dissipation induced non-stationarity in a quantum gas}
\author{Berislav Bu\v ca}
\affiliation{Clarendon Laboratory, University of Oxford, Parks Road, Oxford OX1 3PU, United Kingdom}
\author{Dieter Jaksch}
\affiliation{Clarendon Laboratory, University of Oxford, Parks Road, Oxford OX1 3PU, United Kingdom}
\affiliation{Centre for Quantum Technologies, National University of Singapore, 3 Science Drive 2, Singapore 117543}
\begin{abstract}
Non-stationary long-time dynamics was recently observed in a driven two-component Bose-Einstein condensate coupled to an optical cavity [N. Dogra, et al. arXiv:1901.05974] and analyzed in mean-field theory. We solve the underlying model in the thermodynamic limit and show that this system is always dynamically unstable -- even when mean-field theory predicts stability. Instabilities always occur in higher-order correlation functions leading to squeezing and entanglement induced by cavity dissipation. The dynamics may be understood as the formation of a dissipative time crystal. We use perturbation theory for finite system sizes to confirm the non-stationary behaviour.
\end{abstract}		

\maketitle
\emph{Introduction---} Quantum systems composed of many degrees of freedom are expected to relax to stationarity in the long-time limit. This basic principle has been the subject of intense theoretical and experimental research in recent years, in both equilibrium and non-equilibrium settings and for both open and closed systems, e.g. \cite{Rossini2009,noneq}. Moreover, relaxation to stationarity was shown to happen on relatively short timescales \cite{ETHReview}. In closed systems it is mathematically understood by arguing that every observable evolves as $\ave{O(t)}=\sum_{n,m} e^{\ii \omega_{nm} t} c_{nm}$, where $\omega_{nm}=E_n-E_m$. The coefficients $c_{nm}$ are determined by the initial state and $E_n$ is the eigenvalue of the Hamiltonian $H$ and eigenstate $\ket{E_n}$. For generic observables and initial states the eigenfrequencies $\omega_{nm}$ entering into the time-evolution will be dense and incommensurate. This will lead mutual dephasing (destructive interference) and relaxation to a time-independent (stationary) value of $\ave{O(t)}$ in the long-time limit \cite{Barthel,Esslerreview}.

In contrast, non-stationary dynamics of macroscopic systems is ubiquitous in nature. The emergence of such behavior from the underlying laws of quantum mechanics is an important open question and has far-reaching implications. A possible way for achieving this is for dissipation to dampen all but selected equidistantly spaced frequencies $n \omega_0$ with $n=1,2,3,\cdots$ thus preventing eigenstate thermalization as discussed in \cite{complex} This is to be contrasted with the emergent stationarity due to a dense incommensurate spectrum described above. This mechanism underlies quantum synchronization \cite{qsynch}, dissipative time crystals \cite{timecrystal}, and may induced more complex long-time dynamics \cite{complex} and long-range off-diagonal order \cite{JoeyPaper}. This type of time crystalline behavior is qualitatively different from the standard discrete (or Floquet) time crystal, which is a phase of matter with external time-dependent driving that shows persistent oscillations with a period different from the one of external driving field and may exist both without \cite{discrete1,discrete2} and with \cite{discretdiss1,discretdiss2,discretdiss3,discretdiss4} dissipation. The phenomenon we will be interested in is emergent non-stationarity in model systems without explicit time-dependent external driving (i.e. in a co-rotating frame where the dynamical equations are not explicitly time-dependent), similar to non-stationarity from certain initial states due to many-body scars \cite{scars1}, or the formation of time-crystals in the isolated Heiseberg spin chain \cite{Marko}.   

Long-term non-stationary dynamics was observed in a recent experiment studying a two-component Bose-Einstein condensate (BEC) coherently coupled to two different spatial atomic configurations \cite{Esslinger}, referred to as density mode (DM) and spin mode (SM). The coherent couplings are mediated by photons scattered by the atomic system from a transverse pump field into an optical cavity \cite{note}. The experiment showed a rich phase diagram with a dissipation induced region of instability characterized by persistent oscillations. The system was analyzed in mean-field theory and excellent agreement with the experiment was obtained \cite{note}. 

In this Letter, we provide a solution to the long-time dynamics of the underlying model in the thermodynamic limit by employing the approach pioneered by Emary and Brandes for studying the quantum fluctuations around the mean-field solution \cite{Dicke2}. Our main result is that this system displays non-stationary dynamics for all choices of external parameters. In regions where mean-field theory predicts stability non-stationarity is confined to higher-order correlation functions and counter-intuitively leads to dissipation induced squeezing and entanglement which is not present in the corresponding closed system. We confirm our findings using perturbation theory for a finite system. 

\emph{Two-component BEC coupled to a cavity---}
We model the experimental setup studied in \cite{note} in the Lindblad master equation framework \cite{lindblad,openbook}. By moving to a co-rotating frame we eliminate the (simple) time-dependence coming from the external high-frequency driving and our starting point is a time-independent master equation,
\begin{eqnarray}
&&\frac{\dd}{\dd t}\rho(t) = \LL\rho(t):= \nonumber \\
&&-\ii [H,\rho(t)] +  \kappa \left(2L \rho(t) L^\dagger - \{L^\dagger L,\rho(t)\}\right),
\label{eq:lindeq}
\end{eqnarray}
taking,
\begin{eqnarray}
&&H=\hbar \omega a^\dagger a+ \hbar \omega_0 (J_{\z,+}+J_{\z,-})+\nonumber \\
&&\frac{\hbar}{\sqrt{N}} [\lambda_D(a^\dagger+a)(J_{\x,+}+J_{\x,-})+\nonumber \\
&&\ii \lambda_S(a^\dagger-a)(J_{\x,+}-J_{\x,-})], \label{ham}
\end{eqnarray}
where $a$ ($a^\dagger$) is the annihilation (creation) operator of the cavity mode, $\omega$ is the detuning between the cavity resonance and the transverse pump field, $J_{\alpha,+}$($J_{\alpha,-}$) are the collective spin operators of the $+$($-$) Zeeman state separated by angular frequency $\omega_0$, and $\lambda_{D,S}$ are the coupling strengths of the atomic spins to the cavity mode. The cavity loss is modelled by a single Lindblad operator $L=a$ with rate $\kappa$.

We extend the approach of \cite{Dicke1,Cirac1} to study this system in the thermodynamic limit with the number of particles in the BECs $N \rightarrow \infty$. More specifically, the approach will allow us to study the quantum fluctuations in the leading order of large $N$ around the mean-field solutions. We begin by performing a Holstein-Primakoff transformation,
	\begin{eqnarray}
	J_{+,\pm}&=&b_{1,2}^\dagger \sqrt{N-b_{1,2}^\dagger b_{1,2}},\qquad J_{-,\pm}= J_{+,\pm}^\dagger,
	\label{eq:Hol_Prima}\\
	J_{z,\pm}&=&b_{1,2}^\dagger b_{1,2}-N/2,
	\label{holprim}
	\end{eqnarray}
where $b_1$ ($b_2$) is the bosonic annihilation operator for the $+$ ($-$) BEC. Anticipating instabilities of the ground state with $\ave{J_{z,\pm}}=-N/2$ and $\ave{J_{+,\pm}}=0$ as already obtained in a mean-field treatment \cite{note} we also perform a shift
\begin{equation}
a \to a+\alpha \sqrt{N} \qquad b_{1,2} \to b_{1,2}- \sqrt{\beta_{1,2}} \sqrt{N}. \label{shift}
\end{equation}

We expand the Liouvillian in orders of $N$ keeping only powers higher than $0$. The values of $\alpha, \beta_{1,2}$ are then determined by demanding that the resulting Liouvillian is quadratic. We call $\alpha=\beta_{1,2}=0$ the normal case, and \emph{superradiant} otherwise. Finite values of $\alpha, \beta_{1,2}$ physically correspond to $\ave{ J_{+,\pm}}$ acquiring a non-zero macroscopic mean field value. Our study does not capture the transient build up of the mean-field values $\alpha, \beta_{1,2}$ and thus describes the long-time dynamics. Note that unphysical solutions to $\alpha, \beta_{1,2}$ must be discarded by hand.

In the normal case we obtain for the Hamiltonian (up to an irrelevant constant shift)
\begin{dmath}
 H=\omega a^\dagger a+\omega_0 (b^\dagger_1 b_1+b^\dagger_2 b_2)+\lambda_D(a+a^\dagger)(b_1+b^\dagger_1+b_2+b^\dagger_2 )+\ii \lambda_S (a-a^\dagger)(b_1+b^\dagger_1-b_2-b^\dagger_2 ). \label{Ham_norm}
\end{dmath}
In the superradiant case the Hamiltonian also contains squeezing terms like $b_{1,2}^2$ and $(b^\dagger_{1,2})^2$ and is given in \cite{supp}. The Lindblad operators remain the same following a shift that removes the linear terms in $\LL$. 

Since the resulting Liouvillian is quadratic we solve it exactly using the method of 'third quantization' \cite{3rdquant,3rdquantbos}. The details are in \cite{supp}. 
Expanding the eigenvalues for large $\kappa$ we get in the first two leading orders,
\begin{equation}
\lambda_{n_1,n_2}= \ii n_1  \omega_0 +\frac{2 n_2}{\kappa} \Gamma^2 +\mathcal{O}(\frac{1}{\kappa^2}), \label{eigs}
\end{equation}
where now $n_{1,2}=0,\pm1,\pm2,\dots$. In the normal case $\Gamma^2 = V^2 \equiv \left(\lambda_D^2+\lambda_S^2\right)$ and can be found numerically in the superradiant case.

Eigenvalues with a positive real part are an unphysical consequence of the {\em unbounded} bosonic Liouvillian superoperator. The eigenmodes corresponding to these eigenvalues 'blow up' and signal an instability. This instability arises from the coupling of the lossy cavity to the BECs. We emphasize that the leading imaginary part of the eigenvalues are equally spaced and thus dephasing of the dynamics is prevented \cite{Barthel,Esslerreview}. This is in contrast to the related closed Dicke model, which possesses a dense spectrum and is known to exhibit chaos and thermalization \cite{Dicke2} (see \cite{supp} for more details). It demonstrates how dissipative engineering of the spectrum by coupling to the lossy cavity prevents thermalization. 

Our results are consistent with those obtained in mean-field theory in \cite{note}. Fluctuations around the initial BEC state will be amplified on a time scale given by $\kappa/\Gamma^2$. The system will dynamically evolve away from the initial state and may show persistent oscillations with frequency $\approx \omega_0$. In order to analyse how this instability manifests itself in the dynamics of observables we now move to the Heisenberg picture.

\emph{Equations of motions---} The quadratic Liouvillian admits a finite closed set of Heisenberg equations of motion for $\vec{a}=(a,b_1,b_2,a^\dagger,b^\dagger_1,b^\dagger_2)$ of the form $\dot{\vec{a}}=\LL^\dagger \vec{a}$. It preserves the Gaussian nature of quantum states and hence the system dynamics is fully determined by the one and two point observables. For the one-point functions in the normal case the expectation values evolve according to
 \begin{dmath}
 \left\langle \dot{a}(t)\right\rangle =-\frac{i \lambda _D \left(q_1^*(t)+q_1(t)\right)}{\sqrt{N}}+\frac{\lambda _S \left(q_2(t)^*+q_2(t) \right)}{\sqrt{N}}-\kappa \langle a(t)\rangle -i \omega  \langle a(t)\rangle,
     \end{dmath}
        \begin{dmath}
   \left\langle \dot{b_j}(t)\right\rangle
   =-\frac{i \lambda _D q_0(t)}{\sqrt{N}}+(-1)^{j-1}\frac{\lambda _S q_0'(t)}{\sqrt{N}}-i \omega _0 \left\langle
   b_j(t)\right\rangle,
     \end{dmath}
  where $q_0(t)=\left\langle a^{\dagger }(t)\right\rangle +\langle a(t)\rangle$, $q'_0(t)=\left\langle a^{\dagger }(t)\right\rangle -\langle a(t)\rangle$,  $q_1(t)=\left\langle b_1(t)\right\rangle +\left\langle b_2(t)\right\rangle$, $q_2(t)=\left\langle b_1(t)\right\rangle -\left\langle b_2(t)\right\rangle$. After adiabatic elimination of the cavity mode we obtain agreement with the mean-field treatment in \cite{note}. The superradiant case is treated analogously \cite{supp}.

\begin{figure}[!]
	\begin{center}
		\vspace{0mm}
		\includegraphics[width=0.5\textwidth]{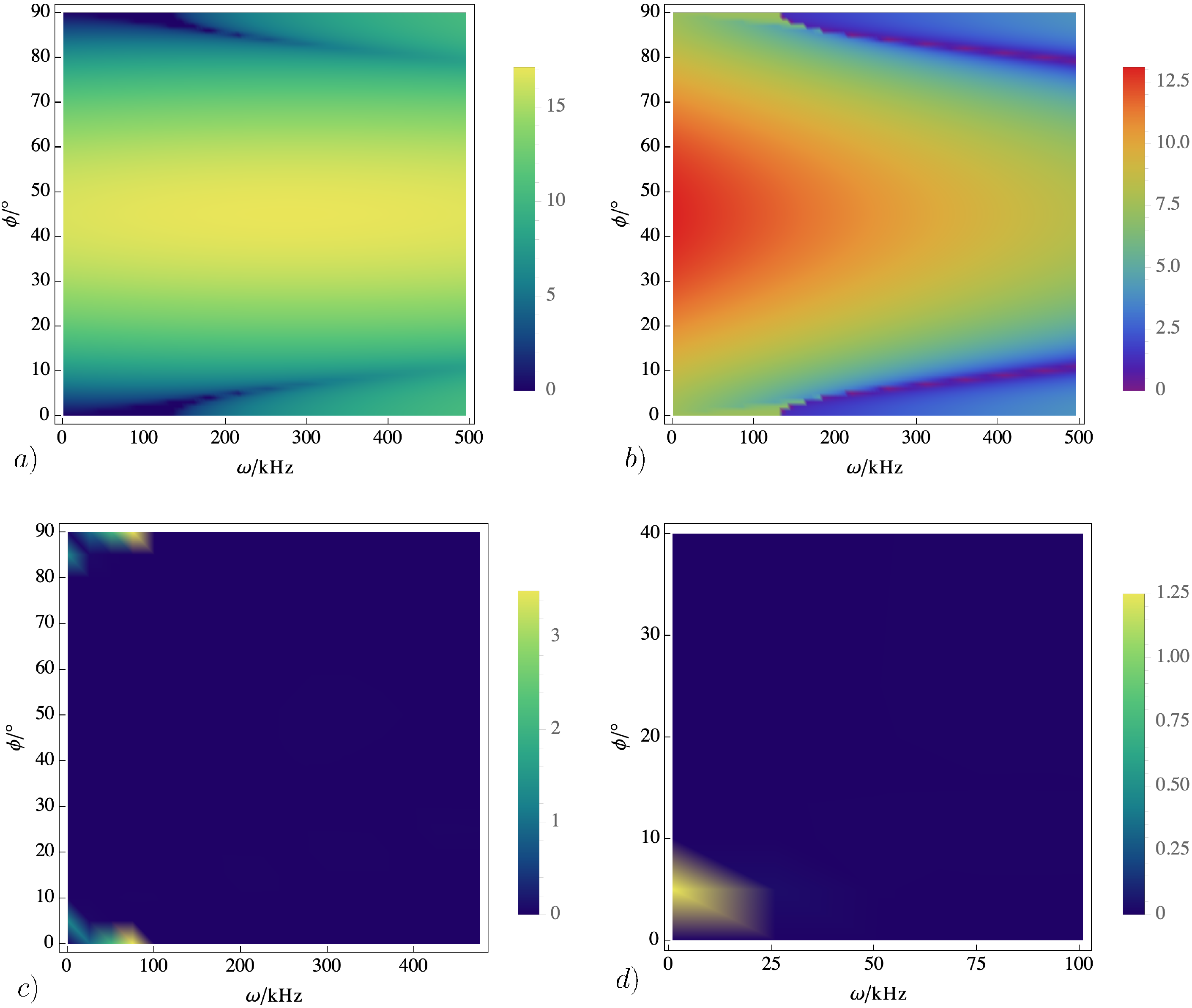}
		\vspace{-5mm}
	\end{center}
	\caption{Stability analysis of the Holstein-Primakoff solutions giving the maximum real part of the eigenvalues time evolution generator for the one-point functions in the normal case a), and the most stable of the superradiant solutions c). In d) we show a non-trivial part of the phase diagram close-up indicating the maximum real part of the most stable of either the normal or superradiant cases. The imaginary part (giving the frequencies of the oscillations) corresponding to a) is shown in b). We use the parametrisation $\lambda_D=V \cos{\phi}, \lambda_S=V \sin{\phi}$ with $V=121.65 \mathrm{kHz}$ and $\omega_0=7.4 \mathrm{kHz}$, $\kappa=1.25 \mathrm{MHz}$.}
	\label{fig:stab}
\end{figure}

We perform a stability analysis of one and two-point correlators in Fig.~\ref{fig:stab} showing the maximum rate at which a fluctuation around the stationary solutions determined by $\alpha$ and $\beta_{1,2}$ can be exponentially amplified. The normal case shown in Fig.~\ref{fig:stab}a) exhibits only a small region around $\phi =\arctan(\lambda_S/\lambda_D)$ where the one-point correlators are stable. This region increases with decreasing strength of the cavity coupling $V$. The instability of the one-point function is accompanied by oscillations with a frequency around $\omega_0$ shown in Fig.~\ref{fig:stab}b) almost everywhere. In contrast, the superradiant case shown in Fig.~\ref{fig:stab}c) is stable for most values of $\phi$ and $\omega$ for the chosen parameters. In Fig.~\ref{fig:stab}d) we see that there is a small part of the phase diagram where neither the normal nor any of the superradiant solutions are stable in the one-point functions. We find that the region of instability around $\phi=90^\circ$ and $\phi=0$ increases with decreasing $V$.

The stability analysis for the two-point correlators gives a different picture. They are also unstable for all of the superradiant solutions except in the points where the model reduces to the Dicke model ($\lambda_D=0$ or $\lambda_S=0$). This is consistent with some of the eigenvalues of the Liouvillian obtained from the rapidities having a positive real part which implies that some of the observables must be unstable.

The instabilities are observable in connected two-point correlation functions of the form $\ave{X_1 X_2}_c:=\ave{X_1 X_2}-\ave{X_1}\ave{X_2}$, where $X_{1,2}$ are BEC observables and we show examples in Fig.~\ref{fig1}. These are related to spin squeezing \cite{spinsq1,spinsq2} (see \cite{supp}). We find that the spin squeezing parameter in the $y$-direction is oscillatory and can be made arbitrarily small with suitable initial choice of $\ave{a^2(0)}$ \cite{supp}, indicating entanglement. The non-trivial behavior of the \emph{connected} correlation function is a clear indication of beyond mean-field behaviour which is of purely quantum origin. The system thus exhibits entanglement induced by dissipation, which could also have ramifications on quantum information processing applications. 

 \begin{figure}[!]
	\begin{center}
		\vspace{0mm}
		\includegraphics[width=0.5\textwidth]{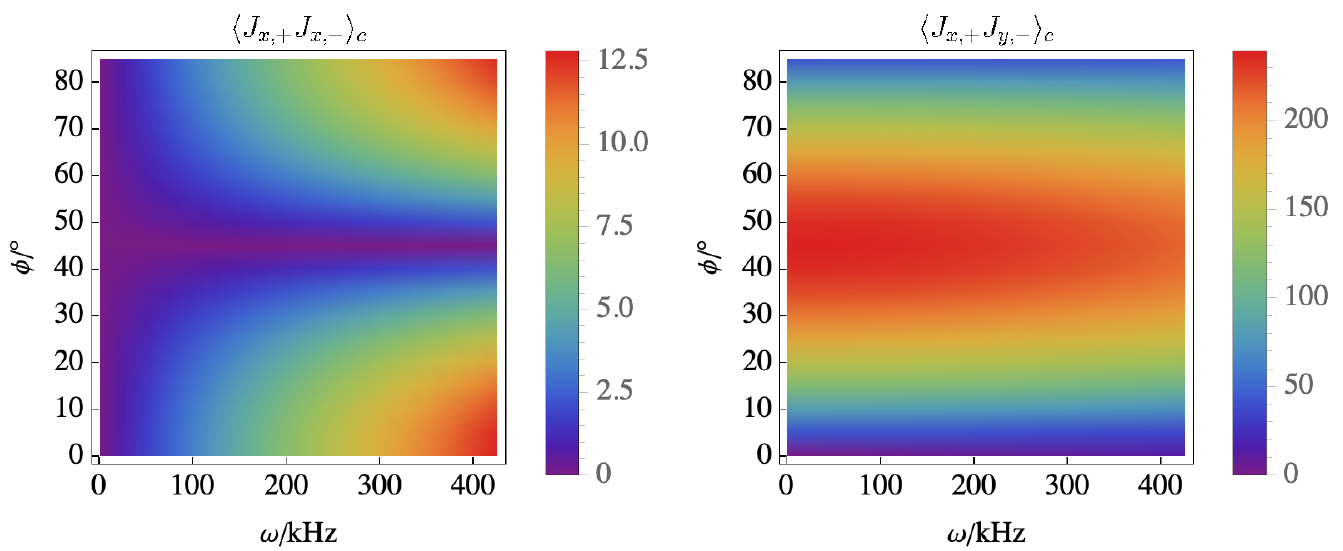}
		\vspace{-10mm}
	\end{center}
	\caption{Examplary maximum values of 2-point functions calculated by time evolving the equations of motions for up to $t=1 \mathrm{ms}$ from the ground state of the two BECs. We plot the amplitude of the oscillations (for the cross-correlators $|\ave{J_{x,+}J_{x,-}}_c|$ and $|\ave{J_{x,+}J_{y,-}}_c|$). We use the parametrisation $\lambda_D=V \cos{\phi}, \lambda_S=V \sin{\phi}$ with $V=92.5\mathrm{kHz}$ and $\omega_0=7.4 \mathrm{kHz}$, $\kappa=1.25 \mathrm{MHz}$. We see clear deviation from mean-field results. Importantly, even deep inside the phase diagram where there are no oscillations of the mean-fields \cite{note}, the cross-correlators  $|\ave{J_{x,+}J_{x,-}}_c|$ and $|\ave{J_{x,+}J_{y,-}}_c|$ do show oscillating behaviour.}
	\label{fig1}
\end{figure}

\begin{figure}[!]
	\begin{center}
		\vspace{0mm}
		\includegraphics[width=0.35\textwidth]{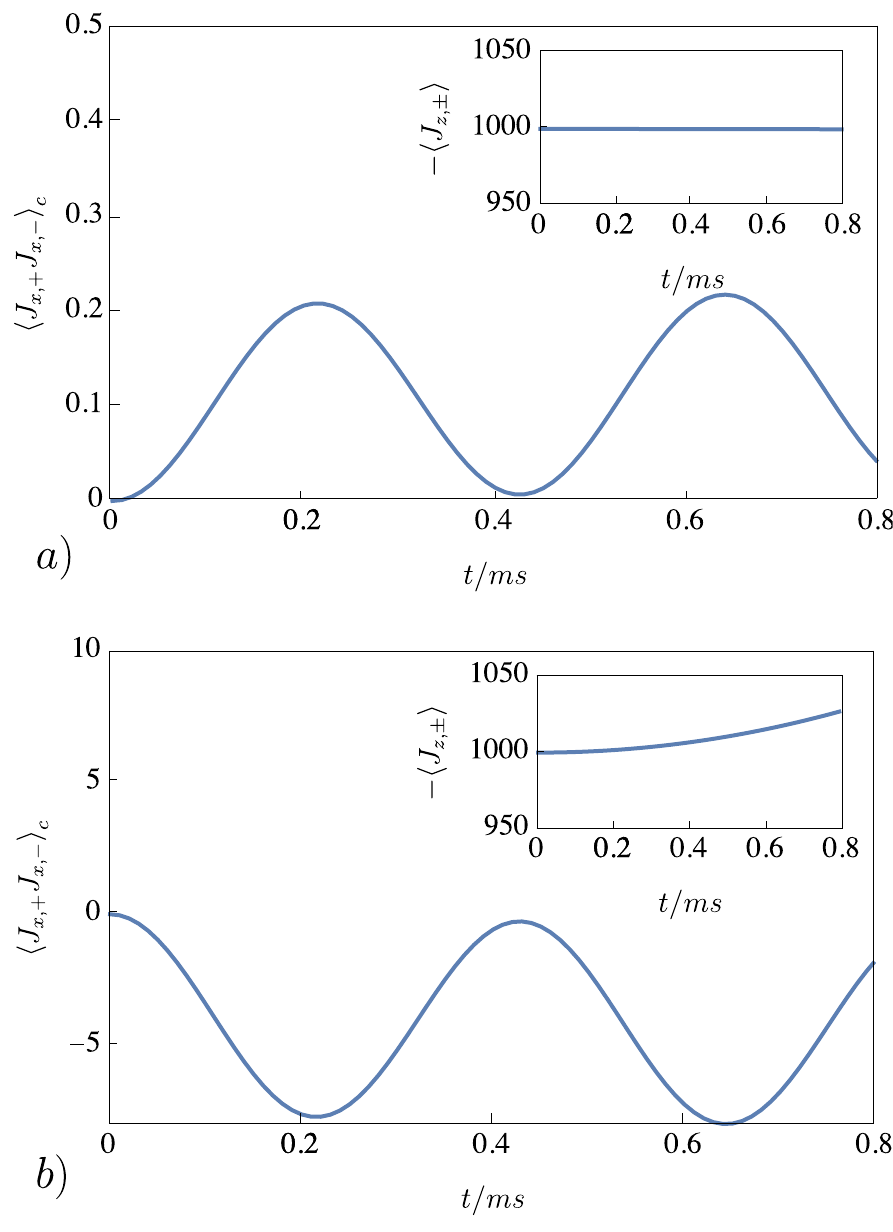}
		\vspace{-7mm}
	\end{center}
	\caption{The cross-correlation functions $\ave{J_{x,+}J_{x,-}}_c=\ave{J_{x,+}J_{x,-}}-\ave{J_{x,+}}\ave{J_{x,-}}$, inside the phase where the one-point functions $\ave{J_{x,\pm}}$ oscillate $\lambda_D=6.3 \mathrm{kHz}, \lambda_S=7.25 \mathrm{kHz},\omega=46 \mathrm{kHz}$ a), and in the phase where they are stationary $\lambda_D=9.6 \mathrm{kHz}, \lambda_S=0.17 \mathrm{kHz},\omega=246 \mathrm{kHz}$ b). The frequency, atom number, and cavity loss are $\omega_0=7.4 \mathrm{kHz}$, $N=2000$ and $\kappa=1.25 \mathrm{MHz}$ in both cases. The insets show that $\ave{J_{z,\pm}}\approx -N/2$ validating the assumptions of \cite{note}.}.
	\label{fig:3}
\end{figure}

We see in Fig.~\ref{fig:3} that even in the phase where the one-point functions relax to stationarity, the cross-correlation $\ave{J_{x,+}J_{x,-}}_c$ and $\ave{J_{x,+}J_{y,-}}_c$ functions are non-stationary. 

Thus the apparent contradiction between mean-field theory showing a phase transition from stable to unstable normal solutions and a full quantum treatment always giving non-decaying eigenvalues with finite imaginary part is resolved: the phase transition takes place for one-point correlators only while higher-order correlations are always unstable. Since there exist eigenmodes of the quantum Liouvillian with eigenvalues that have $\Re(\lambda) \ge0, \im(\lambda)\neq 0$, there will always be some observables for some initial condition that will persistently oscillate.

\emph{Perturbation theory for finite system size ---} The positive real eigenvalues of the Liouvillian in the thermodynamic limit are unphysical and an artefact of the unboundedness of the Liouvillian. Furthermore, since we kept only the linearized quantum fluctuations in the Holstein-Primakoff expansion it is possible that higher order correlators qualitatively modify the main results above. To address these issues we study the system using perturbation theory. We find that the short-time dynamics of \emph{a system of any size} are well-described by the results of the Holstein-Primakoff approach \cite{supp}. More specifically, we obtain eigenfrequencies that are integer multiples of $\omega_0$ in leading order for the longest lived oscillations. The corresponding eigenmodes are givenin \cite{supp} and are perturbatively close to the vacuum state of the cavity.

Using simple large deviation arguments \cite{devi}, it is trivial to argue that the number of photons leaving the cavity should also be low. This is consistent with experimental results in the dynamical instability region (Fig. 2(a) of \cite{note}). Thus we recover the eigenfrequencies of the persistent oscillations, but without the unphysical unbounded increase in the expectation values.

\emph{Conclusion---}
By taking the thermodynamic limit we exactly solved a model of a driven two-component Bose-Einstein condensate coupled to an optical cavity undergoing dissipation \cite{note}. We identified that the system has both normal and superradiant behavior (in which the modes acquire macroscopic mean values) \cite{Dicke1,Dicke2}. We find long-time oscillations in the model due to existence of eigenvalues with non-negative real part and imaginary parts which are close to integer multiples of $\ii \omega_0$. The imaginary part corresponds to the frequencies of the \emph{persistent} oscillations of observables. The fact that these frequencies are almost equidistant means that eigenstate dephasing \cite{Barthel,Esslerreview} of the dynamics is impossible. In contrast, if the cavity were closed, we would have densely spaced and incommensurate frequencies in the Hamiltonian that could mutually dephase and lead to equilibration like in the related Dicke model \cite{Haake1,Haake2}. (see also \cite{limitcycle,semiclassics} for related semiclassical treatment of equilibration).

We thus conclude that we have an example of dissipation induced dynamics in a system that would otherwise equilibrate, akin to a \emph{dissipative time crystal} \cite{timecrystal}. The system in the laboratory frame is time-dependent due to the the (simple) high-frequency driving \cite{note}. Moving to the co-rotating frame leaves us with a time-independent master equation~\eqref{eq:lindeq}. The persistent oscillations arise in in the rotating frame from the interplay between the external drive inputing energy into the system and the cooling from the cavity loss. Without the driving the cavity would simply be empty and without the dissipation the system would heat and thermalize. A sketch of such a mechanism as a possibility for realizing \emph{discrete} time crystals was proposed in \cite{PRX}, but here we have identified an actual physical system with this property. The system we study is better understood in the co-rotating frame as an example of \emph{continuous} time symmetry breaking induced by dissipation as the period of the oscillatory response in the laboratory frame frame bears no fixed relation to the driving period. In other words, the system does not realize a discrete-time crystalline structure in the laboratory frame.

We discussed evidence of spin squeezing in \cite{supp} which should imply entanglement \cite{spinsq2}. We validated the thermodynamic results by taking the large cavity limit and performing perturbation theory for finite systems. We find in the leading order of this perturbation the dark Hamiltonian criteria of Ref.~\cite{complex} are trivially satisfied. The persistent oscillations may thus be also understood as an example of quantum Zeno dynamics (e.g.\cite{Facchi}, \cite{metastability}, \cite{Zanardi}).

In addition to the persistent oscillations at frequencies close to $\omega_0$ the dynamics also entangles the condensates leading to strong squeezing through the dissipative coupling. These features are not captured by mean field theory which instead assumes the condensates to be decoupled. 

In the future we plan to investigate the implications of dissipation induced non-stationary squeezing and entanglement e.g. for quantum enhanced metrology \cite{carlos}.


\emph{Note Added:} While this manuscript was under preparation, the article \cite{andreas} appeared which discusses extensions of the mean-field results of \cite{note}. We go beyond both and study the quantum model.

\emph{Acknowledgments---}  We thank N. Dogra and T. Esslinger for both sharing their preliminary results with us and for useful discussion. We also thank for A. Dietrich and C. Sanchez Munoz for useful discussions. The work here has been supported by EPSRC programme grant EP/P009565/1 and the European Research Council under the European Union's Seventh Framework Programme (FP7/2007-2013)/ERC Grant Agreement No.\ 319286 (Q-MAC)..

	\pagebreak
\onecolumngrid
\begin{center}
	\textbf{\large \emph{Supplemental material}: Dissipation induced non-stationarity in a quantum gas}
\end{center}
\setcounter{equation}{0}
\setcounter{figure}{0}
\setcounter{table}{0}
\setcounter{page}{1}
\makeatletter
\renewcommand{\theequation}{S\arabic{equation}}
\renewcommand{\thefigure}{S\arabic{figure}}
\renewcommand{\bibnumfmt}[1]{[S#1]}
\renewcommand{\citenumfont}[1]{S#1}

\section{Holstein-Primakoff transform}
Let us recall the model studied in the main text. We study the dynamics of the two-component BEC in a driven lossy cavity with the following Lindblad master equation,
\begin{eqnarray}
&&\frac{\dd}{\dd t}\rho(t) = \LL\rho(t)=-\ii [H,\rho(t)] + \kappa \left(2L \rho(t) L^\dagger - \{L^\dagger L,\rho(t)\}\right),
\label{eq:lindeq}
\end{eqnarray}
where,
\begin{eqnarray}
&&H=\hbar \omega a^\dagger a+ \hbar \omega_0 (J_{\z,+}+J_{\z,-})+\frac{\hbar}{\sqrt{N}} [\lambda_D(a^\dagger+a)(J_{\x,+}+J_{\x,-})+\ii \lambda_S(a^\dagger-a)(J_{\x,+}-J_{\x,-})]. \label{ham}
\end{eqnarray}
and the cavity loss is given by a single Lindblad operator $L=a$. The superoperator $\LL$ is the generator of the dynamics. We will be interested in its spectral properties. 
In this section we will study two possible Holstein-Primakoff transformations. 
The standard one, corresponding to the normal phase \cite{SDicke1,SDicke2}, 
	\begin{eqnarray}
	J_{+,\pm}&=&b_{1,2}^\dagger \sqrt{N-b_{1,2}^\dagger b_{1,2}},\qquad J_{-,\pm}= J_{+,\pm}^\dagger,
	\label{eq:Hol_Prima}\\
	J_{z,\pm}&=&b_{1,2}^\dagger b_{1,2}-N/2,
	\label{holprim}
	\end{eqnarray}
gives, after expanding in $1/N$ and keeping only terms up to $N^0$, 
 \begin{dmath}
 H=\omega a^\dagger a+\omega_0 (b^\dagger_1 b_1+b^\dagger_2 b_2)+\lambda_D(a+a^\dagger)(b_1+b^\dagger_1+b_2+b^\dagger_2 )+\ii \lambda_S (a-a^\dagger)(b_1+b^\dagger_1-b_2-b^\dagger_2 ). \label{Ham_norm}
 \end{dmath}
 In certain phases the model may acquire macroscopic stationary values, much like in the related Dicke model \cite{SDicke1,SDicke2}. To treat this we will follow \cite{SDicke1,SDicke2} and perform a shift \emph{prior} to doing the expansion in $1/N$, 
 \begin{equation}
a \to a+\alpha \sqrt{N} \qquad b_{1,2} \to b_{1,2}- \sqrt{\beta_{1,2}} \sqrt{N}. \label{shift}
\end{equation}
Following the Holstein-Primakoff transformation \eqref{holprim} and the shift \eqref{shift} the Hamiltonian becomes, 
 \begin{dmath}
 H=\lambda _D \sqrt{\frac{k_1}{2 N}} \left(\sqrt{N} \alpha ^*+a^{\dagger }+a+\alpha 
   \sqrt{N}\right) \left(-\sqrt{\xi _1} \left(\sqrt{N} \sqrt{\beta _1{}^*}+\sqrt{\beta
   _1} \sqrt{N}\right)+b_1 \sqrt{\xi _1}+\sqrt{\xi _1} b_1{}^{\dagger
   }\right)+\lambda _D \sqrt{\frac{k_2}{2N}} \left(\sqrt{N} \alpha
   ^*+a^{\dagger }+a+\alpha  \sqrt{N}\right) \left(-\sqrt{\xi _2} \left(\sqrt{N}
   \sqrt{\beta _2{}^*}+\sqrt{\beta _2} \sqrt{N}\right)+b_2 \sqrt{\xi _2}+\sqrt{\xi _2}
   b_2{}^{\dagger }\right)+i \sqrt{\frac{k_1}{2N}} \lambda _S
   \left(\sqrt{N} \alpha ^*+a^{\dagger }-a-\alpha  \sqrt{N}\right) \left(-\sqrt{\xi _1}
   \left(\sqrt{N} \sqrt{\beta _1{}^*}+\sqrt{\beta _1} \sqrt{N}\right)+b_1 \sqrt{\xi
   _1}+\sqrt{\xi _1} b_1{}^{\dagger }\right)-i \sqrt{\frac{k_2}{sN}}
   \lambda _S \left(\sqrt{N} \alpha ^*+a^{\dagger }-a-\alpha  \sqrt{N}\right)
   \left(-\sqrt{\xi _2} \left(\sqrt{N} \sqrt{\beta _2{}^*}+\sqrt{\beta _2}
   \sqrt{N}\right)+b_2 \sqrt{\xi _2}+\sqrt{\xi _2} b_2{}^{\dagger
   }\right)+\omega  \left(a \sqrt{N} \alpha ^*+\alpha  N \alpha ^*+\alpha 
   \sqrt{N} a^{\dagger }+a a^{\dagger }\right)+\omega _0 \left(-b_1 \sqrt{N} \sqrt{\beta
   _1{}^*}+\sqrt{\beta _1} N \sqrt{\beta _1{}^*}-\sqrt{\beta _1} \sqrt{N} b_1{}^{\dagger
   }+b_1 b_1{}^{\dagger }-N\right)+\omega _0 \left(-b_2 \sqrt{N} \sqrt{\beta
   _2{}^*}+\sqrt{\beta _2} N \sqrt{\beta _2{}^*}-\sqrt{\beta _2} \sqrt{N} b_2{}^{\dagger
   }+b_2 b_2{}^{\dagger }-N\right), \label{Ham_sup}  \end{dmath}
 where $k_i=2j - \sqrt{\beta^*_i \beta_i}j$, and $\xi_i=\sqrt{1-\frac{b^\dagger_i b_i - \sqrt{\beta_i j}b^\dagger_i - \sqrt{\beta^*_i j}b_i }{k_i} }$. We then expand again in $1/N$ and keep only the zeroth and higher orders. We arrive at a Hamiltonian that contains both quadratic and \emph{linear} terms. Likewise, the Lindblad operators now contain constant terms. We must remove these non-quadratic terms by suitable choices of $\alpha$ and $\beta_i$ so that we can diagonalize the Liouvillian $\LL$ later on \cite{S3rdquantbos}. The corresponding equations for this can be reduced to the following ones, 
 	\begin{align}&\sqrt{\alpha}=\frac{-\lambda _S \left(\sqrt{2-\beta _1 \beta^*_1} \left(\beta^*_1+\beta _1\right)+\left(\beta^*_2+\beta _2\right) \sqrt{2-\beta _2 \beta^*_2}\right)+\ii \lambda
 		_D \left(\sqrt{2-\beta _1 \beta^*_1} \left(\beta^*_1+\beta _1\right)+\left(\beta^*_2+\beta _2\right) \sqrt{2-\beta _2 \beta^*_2}\right)}{\sqrt{2} (\kappa +\ii
 		\omega )} \nonumber\\
 	&\sqrt{\alpha^*}=\frac{-\lambda _S \left(\sqrt{2-\beta _1 \beta^*_1} \left(\beta^*_1+\beta _1\right)+\left(\beta^*_2+\beta _2\right) \sqrt{2-\beta _2 \beta^*_2}\right)+\ii \lambda
 		_D \left(\sqrt{2-\beta _1 \beta^*_1} \left(\beta^*_1+\beta _1\right)+\left(\beta^*_2+\beta _2\right) \sqrt{2-\beta _2 \beta^*_2}\right)}{\sqrt{2} (\kappa -\ii
 		\omega )}\nonumber\\
 	&\frac{\sqrt{2} \alpha \left(3 \beta _1 \beta^*_1+\beta^*_1-4\right) \left(\lambda _S+\ii \lambda _D\right)+i \left(4 \omega _0 \sqrt{2-\beta _1 \beta^*_1}
 		\beta^*_1+\sqrt{2} \alpha^* \left(3 \beta _1 \beta^*_1+\beta^*_1-4\right) \left(\lambda _D+\ii \lambda _S\right)\right)}{\sqrt{\frac{2-\beta _1 \beta^*_1}{N}}}=0\nonumber\\
 	&\frac{\sqrt{2} \alpha  \left(3 \beta _1 \beta^*_1+\beta^*_1-4\right) \left(\lambda _S+i \lambda _D\right)+i \left(4 \omega _0 \sqrt{2-\beta _1 \beta^*_1} \beta^*_1+\sqrt{2} \left(3 \beta _1 \beta^*_1+\beta^*_1-4\right) \alpha ^* \left(\lambda _D+\ii \lambda _S\right)\right)}{\sqrt{\frac{2-\beta _1 \beta^*_1}{N}}}=0\nonumber\\
 	&\frac{\ii \sqrt{2} \alpha  \left(3 \beta _2 \beta^*_2+\beta^*_2-4\right) \left(\lambda _D+i \lambda _S\right)+\sqrt{2} \left(3 \beta _2 \beta^*_2+\beta^*_2-4\right)
 		\alpha ^* \left(\lambda _S+\ii \lambda _D\right)+4 i \omega _0 \sqrt{2-\beta _2 \beta^*_2} \beta^*_2}{\sqrt{\frac{2-\beta _2 \beta^*_2}{N}}}=0\nonumber\\
 	&\frac{\sqrt{2} \alpha  \left(3 \beta _2 \beta^*_2+\beta _2-4\right) \left(\lambda _S-\ii \lambda _D\right)-\ii \left(4 \beta _2 \omega _0 \sqrt{2-\beta _2 \beta^*_2}+\sqrt{2} \left(3 \beta _2
 		\beta^*_2+\beta _2-4\right) \alpha ^* \left(\lambda _D-\ii \lambda _S\right)\right)}{\sqrt{\frac{2-\beta _2 \beta^*_2}{N}}}=0	\label{neweq}	
 	\end{align}
The trivial solution to \eqref{neweq} $\alpha=\beta_i=0$ corresponds to the normal phase from earlier in this section. The other solutions with finite $\beta_i$ correspond to the superradiant phase. After setting $\beta_i$ and $\alpha$ to the non-trivial solutions of \eqref{neweq} the Liouvillian $\LL$ is again quadratic. In the leading order of ${\cal O}(\frac{1}{\kappa^0})$ the solutions to \eqref{neweq} are always $\alpha^{(0)}=\beta^{(0)}_1=\beta^{(0)}_2=0$. This implies that both the normal and superradiant cases coincide in the leading order. In particular, the perturbative results for the eigenfrequencies being integer multiples of $\omega_0$ (purely imaginary leading order contribution to the rapidities) from the main text hold in the leading order of the superradiant case, as well. Note that certain solutions $\alpha,\beta_{1,2}$ may be unphysical, e.g. correspond to expectation values larger than the maximum spin $N$. Such solutions must be discarded by hand.
 
 \section{Bogolioubov transformation for the closed system}
When the system is closed we may perform a simple Bogolioubov transformation and find that the frequencies are given in terms of roots of a third-order polynomial, which read for the normal case,
\begin{dmath}
\frac{1}{4} f^2 \left(-12 \lambda _D^2-12 \lambda _S^2+\omega ^2+2 \omega
   _0^2\right)+\frac{1}{16} f \left[2 \omega _0^2 \left(\omega ^2-6 \left(\lambda
   _D^2+\lambda _S^2\right)\right)-20 \omega  \omega _0 \left(\lambda _D^2+\lambda
   _S^2\right)+36 \left(\lambda _D^2+\lambda _S^2\right){}^2+\omega
   _0^4\right]+\frac{1}{64} \omega _0^2 \left(\omega  \omega _0-2 \left(9 \lambda
   _D^2+\lambda _S^2\right)\right) \left(\omega  \omega _0-2 \left(\lambda _D^2+9
   \lambda _S^2\right)\right)+f^3=0, \label{ens}
\end{dmath}
We find analogous results for the superradiant case upon solving \eqref{neweq}.

 \begin{figure}[h!]
	\begin{center}
		\vspace{0mm}
		\includegraphics[width=0.8\textwidth]{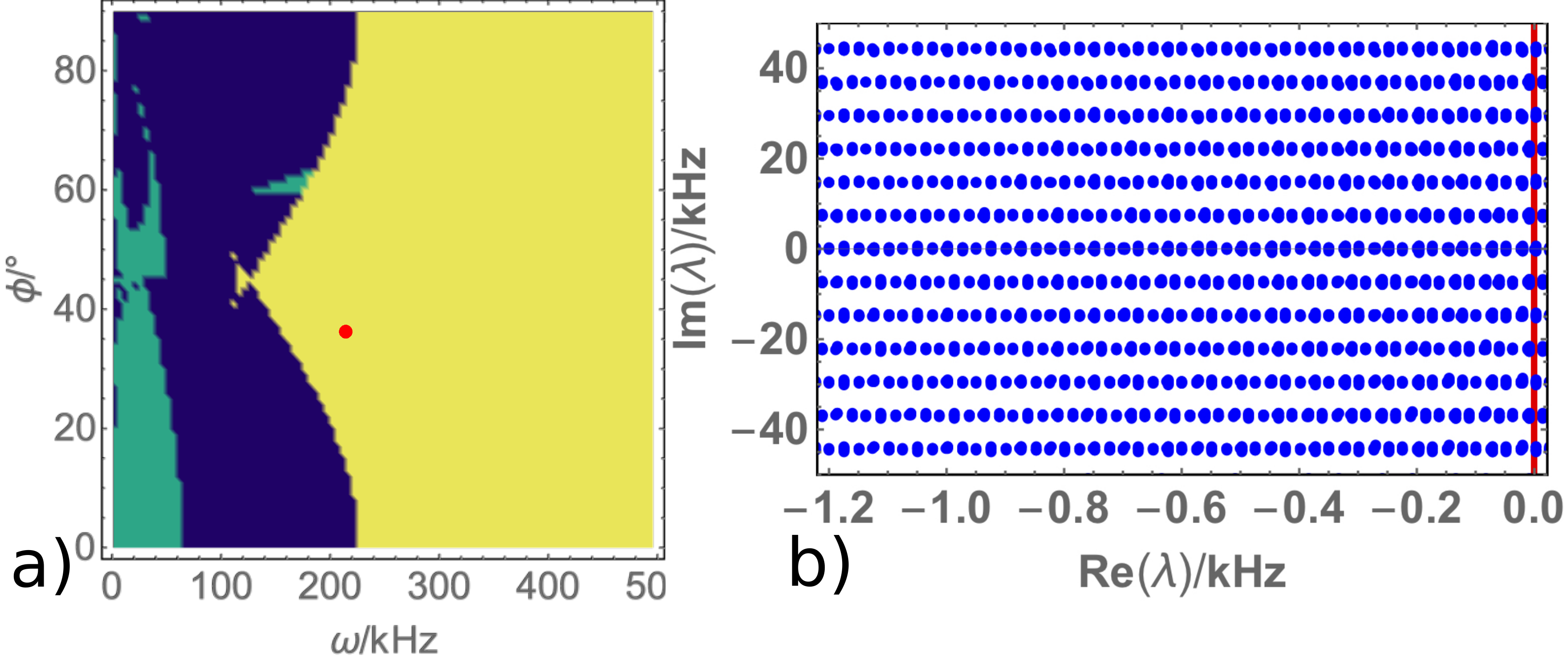}
		\vspace{0mm}
	\end{center}
	\caption{In a) is the phase diagram of the closed model showing the normal phase (yellow), superradiant phase (green) and the phase inaccessible by the Holstein-Primakoff transformation (blue). We use the parametrisation $\lambda_D=V \cos{\phi}, \lambda_S=V \sin{\phi}$ with $V=92.5 \mathrm{kHz}$ and $\omega_0=7.4 \mathrm{kHz}$. In b) we show a part of the real and imaginary parts of the quantum Liouvillian from the red dot in a) for both the closed (red) and open with $\kappa=1.25 \mathrm{MHz}$ case (blue). Note that the spectrum is dense on the imaginary line in the closed case indicating the possibility of eigenstate dephasing. The asymptotic spectrum of the open case features eigenvalues with finite imaginary part and a positive real part. The clusters of eigenvalues in the open case are approximately equally spaced with spacing $\omega_0$.} 
	\label{fig1}
\end{figure}

These roots must be real for the Hamiltonian to be physical. If they are not, we may try applying a shift of the form \eqref{shift} prior to the Holstein-Primakoff transformation. If even following the shift the frequencies are not real, then we cannot treat that phase with the Holstein-Primakoff. This is how we find the phase diagram of the closed model in Fig. 1a. Using these equations we also find the spectrum of the Liouvillian for the closed system and show a part of it in Fig. 1b. More precisely, we solve the equations for the rapidities in both the normal and superradiant phases.

We also show a plot of the spectrum of the quantum Liouvillian in the normal phase for both the closed an open case in Fig. 1b. The closed case has a dense set of frequencies on the imaginary axis, whereas the open one does not. This demonstrates the discussion in the Introduction of the main text and presents the basic idea of dissipative spectral engineering. 

\section{Third quantization}
In this section we briefly recall the method of third quantization and apply it to our problem. We follow the work of Prosen \cite{S3rdquant}. More specifically, we will now recall the form of third quantization applied to bosons given by Prosen and Seligman \cite{S3rdquantbos}. 

\begin{figure}[!]
	\begin{center}
		\vspace{0mm}
		\includegraphics[width=0.5\textwidth]{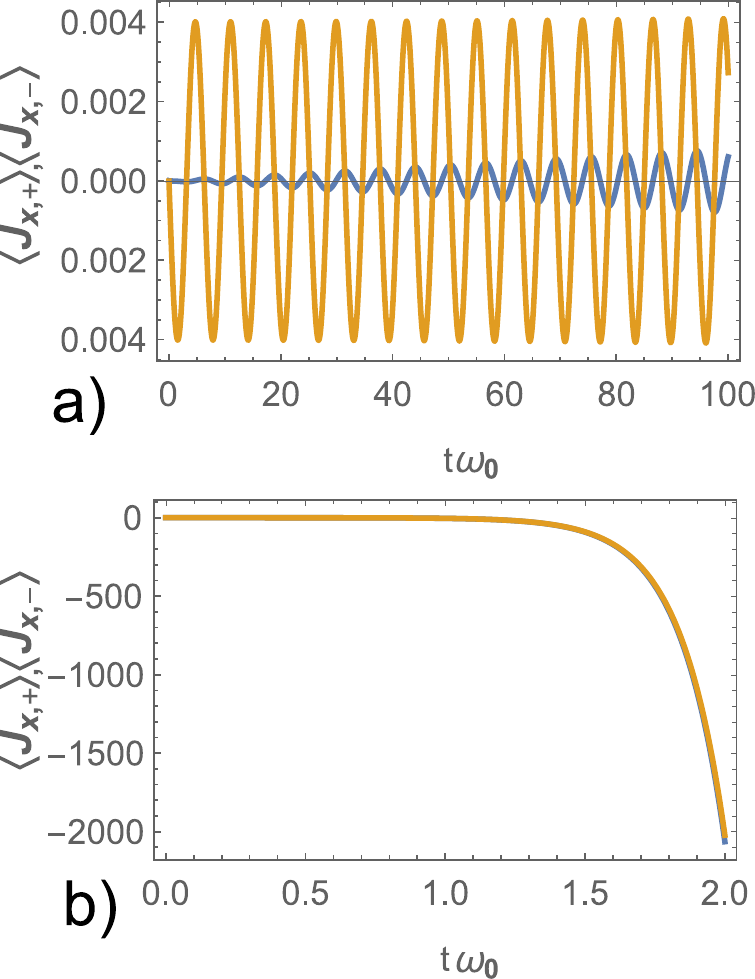}
		\vspace{-5mm}
	\end{center}
	\caption{The induction of oscillations of $J_{x,+}$ (blue) and $J_{x,-}$ (orange) in the thermodynamic limit in the normal phase. We set $\kappa=1000 \hbar \omega$, and a) $\lambda_S=\lambda_D=\omega_0$, b) $\lambda_S=\omega_0,\lambda_D=100\omega_0$, $\omega=1000 \omega_0$,. Note that the frequencies of the oscillations are close to integer multiplies of $\omega_0$. In b) the BECs do not oscillate, rather they acquire diverging stationary values.}
	\label{figsup}
\end{figure}

For sake of brevity we use vector notation $\underline{a}:=(a_1,a_2,a_3)=(a,b_1,b_2)$. Superoperators will be denoted with a hat $\hat{O}$. We also define left and right multiplication superoperators, $\hat{O}^L x:=O x$ and $\hat{O}^R x:=x O$. We now define the following superoperator maps, 
\begin{align}
&\hat{a}_{0,j}=\hat{a}^L_j, \qquad \hat{a}'_{0,j}=\hat{a^\dagger}^L_j -\hat{a^\dagger}^R_j, \nonumber \\
&\hat{a}_{1,j}=\hat{a^\dagger}^R_j, \qquad \hat{a}'_{1,j}=\hat{a}^R_j -\hat{a}^L_j. \label{supop}
\end{align}
Crucially, these operator satisfy almost-canonical commutation relations,
\begin{equation}
[\hat{a}_{\nu,j}\hat{a}'_{\mu,k}]=\delta_{\nu,\mu}\delta_{j,k} \qquad [\hat{a}_{\nu,j}\hat{a}_{\mu,k}]=[\hat{a}'_{\nu,j}\hat{a}'_{\mu,k}]=0.
\end{equation}
We write the Hamiltonian as,
\begin{equation}
H=\underline{a}^\dagger\cdot {\bf H}\cdot\underline{a}+ \underline{a}\cdot{\bf K}\cdot \underline{a}+ \underline{a}^\dagger\cdot{\bf K}^*\cdot \underline{a}^\dagger+\underline{f} \cdot\underline{a}+\underline{f}^*\cdot\underline{a}^\dagger, \label{3ham}
\end{equation}
where ${\bf H}$ and ${\bf K}$ are $6 \times 6$ matrices acting on the space of $(a_1,a_2,a_3,a^\dagger_1,a^\dagger_2,a^\dagger_3)$, and $\underline{f}$ is a 6-dimensional vector. These matrices and this vector can be easily read out of \eqref{Ham_norm} (or \eqref{Ham_sup} for the shifted case). We write the Lindblad operator as,
\begin{equation}
L=\underline{l} \cdot \underline{a}+\lambda, \label{3lind} 
\end{equation}
where $\underline{l}=(1,0,0,0,0,0)$ and $\lambda$ is a constant shift coming from \eqref{shift}. The full Liouvillian is then, 
\begin{align}
&\LL=-\ii \hat{H}^L+\ii \hat{H}^R+\kappa \left( 2\hat{L^{}}^L\hat{L^\dagger}^R-\hat{L^\dagger}^L\hat{L^{}}^L -\hat{L^{}}^R\hat{L^\dagger}^R \right) \nonumber\\
&=-\ii \underline{\hat{a}}'_0 \cdot {\bf H} \cdot \underline{\hat{a}}_0 +\ii \underline{\hat{a}}'_1 \cdot {\bf H}^* \cdot \underline{\hat{a}}_1+\ii  \underline{\hat{a}}'_1 \cdot {\bf K} \cdot(2\underline{\hat{a}}_0 +\underline{\hat{a}}'_1)-\ii  \underline{\hat{a}}'_0 \cdot {\bf K}^* \cdot(2\underline{\hat{a}}_1 +\underline{\hat{a}}'_0)+\underline{\hat{a}}'_0 \cdot {\bf M}\cdot \underline{\hat{a}}_0-\underline{\hat{a}}'_1 \cdot {\bf M}\cdot \underline{\hat{a}}_1+ \underline{g} \cdot \underline{a}',
\end{align}
 where $\underline{\hat{a}}_{0,1}$($\underline{\hat{a}}'_{0,1}$) are the corresponding vectors whose components are defined in \eqref{supop}, ${\bf M}=\kappa (l \otimes l)$, and $\underline{g}$ is a linear shift coming from $\underline{f}$ and $\lambda$  \eqref{3ham}, \eqref{3lind}. In order to remove this linear term and make the Liouvillian fully quadratic we will need to eliminate $\underline{g}$. This is done via solving \eqref{neweq}. Following this the Liouvillian can be written as, 
 \begin{equation}
 \LL=\underline{\hat{b}} \cdot {\bf S}\cdot \underline{\hat{b}}-S_0 \one,
 \end{equation}
where $\underline{b}$ is a \emph{12-dimensional} vector $\underline{b}:=(\hat{a}_0,\hat{a}_1,\hat{a}'_0,\hat{a}'_1)$, and,
\begin{equation}
{\bf S}= \begin{pmatrix}
  {\bf 0} & -{\bf X} \\
  -{\bf X}^T  & {\bf Y}
 \end{pmatrix},
\end{equation}
 where we further have,
 \begin{equation}
 {\bf X}:=\frac{1}{2} \begin{pmatrix}
  \ii {\bf H}^*+{\bf M} & -2\ii {\bf K} \\
  2\ii {\bf K}^*  &  -\ii {\bf H}+{\bf M}^*
 \end{pmatrix},
 \end{equation}
 and,
  \begin{equation}
 {\bf Y}:=\frac{1}{2} \begin{pmatrix}
  -2\ii {\bf K}^* &{\bf 0} \\
 {\bf 0}  &  2\ii {\bf K}
 \end{pmatrix}.
 \end{equation}
 The scalar $S_0=\tr {\bf M}$. Assuming that ${\bf X}$ is diagonalizable, 
 \begin{equation}
 {\bf X}={\bf P} {\bf \Delta} {\bf P}, \qquad {\bf \Delta}={\rm diag}(\chi_1,\chi_2,\ldots,\chi_6),
 \end{equation}
 where $\chi_j$ we call the \emph{rapidities}. It can be shown that diagonalizing ${\bf X}$ and solving the following matrix equation,
 \begin{equation}
 {\bf X}^T \cdot {\bf Z}+{\bf Z} \cdot {\bf X}={\bf Y},
 \end{equation}
 for ${\bf Z}$ provides the full diagonalization of $\LL$ \cite{S3rdquantbos} . Namely, the Liouvillian can be written in terms of \emph{normal master modes},
 \begin{equation}
 \underline{\hat{\zeta}}={\bf P}^T \cdot(\underline{\hat{a}}-{\bf Z}\underline{\hat{a}}'), \qquad  \underline{\hat{\zeta}}'={\bf P}^{-1} \cdot(\underline{\hat{a}}'),
 \end{equation}
as,
\begin{equation}
\LL=-2 \sum_{k}^{6} \chi_k \hat{\zeta}'_k\hat{\zeta}_k. 
\end{equation}
As the normal modes satisfy almost-canonical commutation relations, 
\begin{equation}
[\hat{\zeta}_k,\hat{\zeta}'_j]=\delta_{k,j}, \qquad [\hat{\zeta}_k,\hat{\zeta}_j]= [\hat{\zeta}'_k,\hat{\zeta}'_j]=0,
\end{equation}
we may construct the entire spectrum of $\LL$ from the stationary state $\rho_\infty$ ($\LL \rho_\infty=0$),
\begin{equation}
\LL \rho_k=\lambda_k \rho_k, \qquad \rho_k=\prod_{r}\frac{\left(\hat{\zeta}_r'\right)^{n_r}}{\sqrt{n_r!}}\rho_\infty, \qquad \lambda_k=-2\sum_{r} n_r \chi_r,
\end{equation}
where we have slightly abused notation by indexing with $k$ all the possible values of $n_r=0,1,2,\ldots$. This follows from the fact that $\hat{\zeta}'_r$ function as a raising superoperators for $\LL$, $[\LL,\hat{\zeta}_r']=\chi_r \hat{\zeta}_r'$. The stationary state itself is Gaussian and can be calculated from the 2-point correlation functions only, 
\begin{equation}
\tr \hat{a}_r \hat{a}_s \rho_\infty=Z_{r,s}.
\end{equation}
In the normal phase the rapidities $\chi_k$ are given in terms of roots of the following 6th order polynomial,
\begin{dmath}
16 \chi ^4 \left(-12 \lambda _D^2+\kappa ^2-12 \lambda _S^2+\omega ^2+2 \omega
   _0^2\right)+4 \chi ^2 \left[2 \omega _0^2 \left(-6 \left(\lambda _D^2+\lambda
   _S^2\right)+\kappa ^2+\omega ^2\right)-20 \omega  \omega _0 \left(\lambda
   _D^2+\lambda _S^2\right)+36 \left(\lambda _D^2+\lambda _S^2\right){}^2+\omega
   _0^4\right]+\omega _0^2 \left[-20 \omega  \omega _0 \left(\lambda _D^2+\lambda
   _S^2\right)+4 \left(9 \lambda _D^2+\lambda _S^2\right) \left(\lambda _D^2+9 \lambda
   _S^2\right)+\omega _0^2 \left(\kappa ^2+\omega ^2\right)\right]+\chi ^3 \left(96
   \kappa  \left(\lambda _D^2+\lambda _S^2\right)-32 \kappa  \omega _0^2\right)+\chi 
   \left(24 \kappa  \omega _0^2 \left(\lambda _D^2+\lambda _S^2\right)-4 \kappa 
   \omega _0^4\right)-64 \kappa  \chi ^5+64 \chi ^6=0. \label{normrap}
\end{dmath}
Then expanding for large $\kappa$ we arrive to the simple result of the main text for the eigenvalues $\lambda$. Analogous equations for the rapidities may be obtained in the superradiant phase. We show in Fig.~\ref{figsup} how the mean field found in \cite{Snote} forms from the quantum fluctuations. 

It is crucial to note that since the equations of motion for the observables are linear for all one and two-point correlators choosing initial values of all observables to be 0 will lead to them staying 0 for all time. It is for this reason that as the initial state we must pick a non-zero value for $\ave{a(0)}$, which is physically reasonable in this driven system. We find, however, by integrating the differential equations numerically that the precise value of $\ave{a(0)}$ only determines the rate of divergence of the unstable observables. We also find that by varying $\ave{a^2(0)}$ we can make the spin squeezing \cite{Sspinsq1,Sspinsq2} in the y-direction $\xi_{y,\pm}=\frac{N (\Delta J_{y,\pm})^2}{\ave{J_{z,\pm}^2}+\ave{J_{x,\pm}^2}}$ arbitrarily small.  

\section{Semiclassical limit}

Here we give the semiclassical approximation of the master equation from the main text. We start from the Heisenberg picture for $\ave{a(t)}=\alpha,\ave{J_{-\pm}(t)}=\beta_{1,2},\ave{J_{z,\pm}(t)}=w_{1,2}$, which we assume factor $\ave{O_1 O_2}=\ave{O_1}\ave{O_2}$ (noting that $\ave{O^\dagger}=\ave{O}^*$),
\begin{align}
&\dot{\alpha}=\alpha (-\kappa -i \omega )-i\lambda_D
   \left(\beta_1^*+\beta_2^*+\beta_1+\beta_2\right)
   +\lambda_S
   \left(\beta_1^*-\beta_2^*+\beta_1-\beta_2\right),\\
&\dot{w}_1'=-\lambda_S \left(\alpha^*-\alpha\right)
   \left(\beta_1-\beta_1^*\right)+i\lambda_D
   \left(\alpha^*+\alpha\right)
   \left(\beta_1-\beta_1^*\right),\\
&\dot{w}_2'=\lambda_S \left(\alpha^*-\alpha\right)
   \left(\beta_2-\beta_2^*\right)+i\lambda_D
   \left(\alpha^*+\alpha\right)
   \left(\beta_2-\beta_2^*\right),\\
   &\dot{\beta}_1=2 i
  \lambda_D w_1 \left(\alpha^*+\alpha\right)-2
   \lambda_S w_2 \left(\alpha^*-\alpha\right)-i
   \omega_0 \beta_1,\\
   &\dot{\beta}_2=2 i\lambda_D
   w_2 \left(\alpha^*+\alpha\right)+2 \lambda_S
   w_2 \left(\alpha^*-\alpha\right)-i \omega_0
   \beta_2
\end{align}
The crucial difference to the mean-field approximation used in \cite{Snote} is that the semiclassical equations allow for dynamics in the $\ave{J_{z,\pm}}$, too. 

\section{Perturbation theory for finite system size} 

As in the main text we take large $\kappa$, i.e., $\kappa=\gamma \kappa'$ and that $\gamma \gg 1$, $\kappa' \approx \omega_0$.  This allows us to do perturbation theory.
We split $\LL$ into $\LL=\gamma(\LL^{(0)}+ 1/\gamma \LL^{(1)})$,
where,
\begin{eqnarray}
&&\LL^{(0)} \rho=-\ii [H^{(0)},\rho] +  \kappa'\left(2L \rho L^\dagger - \{L^\dagger L,\rho\}\right), \quad H^{(0)}=0,\nonumber\\
&&\LL^{(1)} \rho=-\ii [H,\rho]\label{eq:lindeq1}.
\label{eq:lindeq0}
\end{eqnarray}
We are interested in both the right and left eigenvectors of $\LL$,
\begin{equation}
\LL \rho=\lambda \rho \qquad \sigma \LL=\lambda \sigma. \label{eigeqs}
\end{equation}
We then formally expand $\rho=\rho^{(0)}+1/\gamma \rho^{(1)}+1/\gamma^2 \rho^{(2)}+{\cal O}(1/\gamma^3)$ and likewise for $\lambda$ and $\sigma$. We insert this into the eigenvalue equations \eqref{eigeqs} and collect the same orders in $\gamma$. The leading order for $\lambda^{(0)}=0$, 
$$\LL^{(0)} \rho^{(0)}_{\infty,n_\pm,m_\pm}=0$$,
\begin{equation}
\rho^{(0)}_{\infty,n_\pm,m_\pm}=A_{+}^{n_+}A_{-}^{n_-}\ket{vac}_c\bra{vac}_c\otimes\ket{0,0}\bra{0,0}(A_{+}^{m_+}A_{-}^{m_-})^{\dagger}, \label{leading}  \end{equation}
where $A_{\pm}=\one_c \otimes J_{+,\pm}$, and $J_{+,\pm}:=1/2(J_{x,\pm}+\ii J_{y,\pm})$ (with the subscript $c$ denoting the cavity part of the Hilbert space). The NESS subspace is highly degenerate, as the $A_{\pm}$ operators \cite{Scomplex} trivially commute with both $L$ and $H^{(0)}$ \cite{SBucaProsen}. The left stationary states are solved similarly with,
\begin{equation}
\sigma^{(0)}_{\infty,n_\pm,m_\pm}=A_{+}^{n_+}A_{-}^{n_-}\left(\one_c\otimes\ket{0,0}\bra{0,0}\right)(A_{+}^{m_+}A_{-}^{m_-})^{\dagger}.  \end{equation}

In the next order we find that the degenerate eigenvalue 0 is split in analogy with standard perturbation theory. By acting with $(\sigma^{(0)}_{\infty,n_\pm,m_\pm})^\dagger$ from the left and using biorthogonality  $\tr (\sigma^\dagger_j \rho_k)=\delta_{j,k}$.,
\begin{equation}
\lambda^{(1)}_{n_\pm,m_\pm,n'_\pm,m'_\pm}= -\ii \tr \left(\sigma^{(0)}_{\infty,n'_\pm,m'_\pm} [H^{(1)},\rho^{(0)}_{\infty,n_\pm,m_\pm}]\right).
\end{equation}
This may be easily evaluated and gives that the longest lived oscillating observables will oscillate with eigenfrequencies that are integer multiples of $\omega_0$ in the leading order.    The fact that if the perturbation is purely Hamiltonian, the eigenvalues coming from the splitting of eigenvalue 0 of $\LL$ are purely imagainary has been appreciated before, e.g. \cite{Smetastability,SZanardi}. The real correction to these imaginary eigenvalues will be of order $1/\gamma$ smaller than the imaginary part. The corresponding eigenmodes are given by \eqref{leading} and are perturbatively close to the vacuum state in the cavity.

We are also interested in the statistics of the number of photons leaving the cavity, which is another experimentally relevant parameter \cite{Snote}. To compute this we apply the method of large deviations \cite{Sdevi}. Mathematically, this corresponds to introducing a \emph{counting field} $\chi$ in the Liouvillian that 'counts' how much photons enter ($e^{+ \ii \chi}$), or leave ($e^{-\ii \chi}$) the cavity. The cumulant generating function for the flow of photons will be given by the leading eigenvalue of this 'deformed' Liouvillian \cite{Sdevi}.

In the leading order, in $\LL^{(0)}$ the only term changing the number of photons is $L$. It is straightforward to show that the counting field does not change the eigenspectrum of $\LL^{(0)}$, meaning that there is no flow of photons in the leading order of strong cavity loss.

Both the result for the frequencies of the oscillations and the number of photons leaving the cavity are consistent with the experimental observation in the dynamical instability region (Fig. 2 of \cite{note}) where we may assume that $\kappa \gg \omega_0,\lambda_d,\lambda_S $

\end{document}